\documentclass[mathleft
]{an}
\usepackage{graphicx}
\usepackage{times}
\usepackage{amssymb}
\begin{document}

\Pagespan{789}{}
\Yearpublication{2006}%
\Yearsubmission{2005}%
\Month{11}%
\Volume{999}%
\Issue{88}%

\title{Unveiling the X-ray/UV properties of disk winds in active galactic nuclei using broad
and mini-broad absorption line quasars}

\author{M. Giustini\thanks{Corresponding author
  \email{m.giustini@sron.nl}\newline}
}
\titlerunning{X-ray/UV properties of AGN winds using BAL and mini-BAL QSOs}
\authorrunning{M. Giustini}
\institute{
SRON - Netherlands Institute for Space Research, Sorbonnelaan 2, 3584 CA Utrecht, Netherlands
}

\received{ }
\accepted{ }
\publonline{later}


\abstract{
We present the results of the uniform analysis of $46$ XMM-Newton observations of six BAL and seven mini-BAL QSOs belonging to the Palomar-Green Quasar catalogue.
Moderate-quality X-ray spectroscopy was performed with the EPIC-pn, and allowed to characterise the general source spectral shape to be complex, significantly deviating from a power law emission. A simple power law analysis in different energy bands strongly suggests absorption to be more significant than reflection in shaping the spectra.
If allowing for the absorbing gas to be either partially covering the continuum emission source or to be ionised, large column densities of the order of $10^{22-24}$ cm$^{-2}$ are inferred. When the statistics was high enough, virtually every source was found to vary in spectral shape on various time scales, from years to hours. All in all these observational results are compatible with radiation driven accretion disk winds shaping the spectra of these intriguing cosmic sources.
 }

\maketitle

\section{Introduction}

In recent years, winds were revealed as a fundamental ingredient of Active Galactic Nuclei (AGN).
Broad absorption line quasars (BAL QSOs) are AGN displaying in their optical/UV spectra broad absorption troughs due to resonant transitions of highly ionised metals such as Si IV, C IV, N V, O VI, strongly blueshifted (from a few $1,000$ km s$^{-1}$ up to several $10,000$ km s$^{-1}$) with respect to the host galaxy systemic redshift, indicating the presence of powerful nuclear outflows (e.g. Weymann, Carswell \& Smith 1981).   

BALs are quite common: they are observed in about $10-15$\% of optically selected AGN (e.g. Weymann 1991), and their intrinsic fraction is estimated to be about $20-25$\% (e.g. Trump et al. 2006; Knigge et al. 2008; Gibson et al. 2009). 
Despite being known for almost 50 years (\textit{``The spectrum of the quasi-stellar source PHL 5200 appears to be more interesting  than that of any source studied thus far at Kitt Peak''}, Lynds 1967), their physical nature is still elusive. 
Is the BAL QSO a phase in the lifetime of the AGN lasting $\sim 20-25$\% of its life? Or is the outflowing gas responsible for the UV BALs present in every AGN, but covering only $\sim 20-25$\% of the sky as seen by the source? 
In other words: is the BAL QSO appearance a temporal effect depending on the phase of the QSO life (e.g. Voit, Weymann \& Korista 1993; Farrah et al. 2007), or is it a geometrical effect depending on our line of sight as in accretion disk wind scenarios (e.g. Murray et al. 1995)? 
All in all, BAL QSOs are extremely interesting sources, as they are potential direct probes of \textit{both} the evolution and the geometry of AGN, helping unveiling both the physics of accretion/ejection around supermassive black holes (SMBHs), and the interplay between the SMBH and its environment during the AGN phase. 

Powerful winds can be launched and accelerated in accretion disks around accreting SMBHs by three main mechanisms: thermal, radiation, or magnetic pressure. 
Simple thermal winds are able to explain the presence of very low-velocity winds (with outflow up to a few $1,000$ km s$^{-1}$) in AGN, but in order to explain the commonly observed high-velocities (up to $0.2c$) either radiation-driven or magnetically-driven winds must be invoked (see e.g. Proga 2007; Everett 2007). 
Murray et al. (1995) were the first to successfully apply the UV line-driving mechanism to AGN accretion disk models, and to demonstrate that it would be able to accelerate powerful winds with velocities and geometry comparable with the observations. In their model, the UV-absorbing wind needed to be shielded from the strong ionising continuum by a thick layer of gas (the so-called ``hitch-hiking gas''), able to absorb the high-energy continuum emission and thus preventing the UV-absorbing wind from becoming overionised and therefore losing any capability to be pushed by resonant line pressure. 
Hydrodynamical simulations of UV line-driven accretion disk winds show how a layer of thick and cold absorbing gas naturally arises in the inner regions of the accretion/ejection flow (the ``failed wind'', Proga, Stone \& Kallman 2000; Proga \& Kallman 2004).
The geometry and the dynamics of radiation-driven accretion disk winds depend critically on the AGN UV/X-ray flux ratio and on the mass accretion rate.
Conversely, in magnetically-driven scenarios there is no need for a X-ray absorbing shield in order to launch and accelerate powerful accretion disk winds (e.g. Konigl \& Kartije 1994; Everett 2005); the ionisation state of the wind will be however still dependent on the presence of an X-ray shield (Fukumura et al. 2010).  

X-ray/UV observations are therefore crucial to distinguish between the two main scenarios for AGN accretion disk winds, thus helping clarify the physics of the inner regions of AGN and the impact in terms of energetics of such winds in the host galaxy environment.
X-ray observations of BAL QSOs are however extremely challenging: one one hand, the vast majority of known BAL QSOs has been selected in ground-based surveys, and lie at high redshift (the most common observed CIV $\lambda 1549$ transition enters the visible band for sources at $z\gtrsim 1.7$); one another hand, BAL QSOs are historically known to be X-ray weak: at a given UV luminosity, their X-ray flux is typically $10-30$ times lower than a non-BAL QSO (Green et al. 1995; Laor et al. 1997).

Such X-ray weakness seems consistent with be due to X-ray absorption (Brandt, Laor \& Wills 2000; Gallagher et al. 2006), and with BAL QSOs having an underlying typical spectral energy distribution, supporting again geometrical scenarios. 
The large ($N_H\sim 10^{23-24}$ cm$^{-2}$) X-ray absorbing column densities inferred from low S/N X-ray observations seem to favour the radiation-driven scenario; however, a sample of X-ray bright BAL QSOs has been described by Giustini, Cappi \& Vignali (2008) and shown to be much less absorbed than classical X-ray weak BAL QSOs. 
At the same time, it was shown how the strength of the UV absorption troughs in BAL QSOs strongly correlates with the amount of cold X-ray absorption, and the highest velocity, highest equivalent width BALs happen in QSOs that are most likely Compton-thick. 
Strong relativistic effects such as light bending have also been invoked as possible explanation for the observed X-ray weakness of BAL QSOs (e.g. Schartel et al. 2005), and given the moderate quality of the available X-ray spectra, often strong absorption and strong gravity effects models were statistically equivalent (e.g. Ballo et al. 2008).
Recently, thanks to the large bandwidth of NuSTAR, a significant population (about one third of the total according to the first estimates) of BAL QSOs that are intrinsically X-ray weak has emerged (Luo et al. 2013; 2014). 

Historically, BALs are defined to have a measured FWHM $> 2,000$ km s$^{-1}$, in order to be sure not to include spurious non-intrinsic-to-the-QSO (i.e. intervening) absorbers within low signal-to-noise ratio spectroscopic surveys.
 Other intrinsic blueshifted absorbers, narrower than the BALs, but involving the same ionic transitions and the same outflow velocity range, are often observed in the UV band of AGN. 
 Depending on the width of the absorption troughs, quasars hosting such features are classified as mini-Broad Absorption Line Quasars (mini-BAL QSOs, $500$ km s$^{-1}$  $<$ FWHM $< 2,000$ km s$^{-1}$), and Narrow Absorption Line Quasars (NAL QSOs, FWHM $< 500$ km s$^{-1}$); see Hamann \& Sabra 2004 for a review. 
 NAL and mini-BAL QSOs are much less studied than BAL QSOs, and they are estimated to be present in $40-50$\% and $5-15$\% of AGN, respectively (Ganguly et al. 2001, Rodr\'iguez-Hidalgo et al. 2007; Ganguly \& Brotherton 2008; Hamann et al. 2012).
The physical relation between these UV absorbers with different widths is far from being established; observationally, mini-BAL QSOs have properties in between those of BAL and non-BAL QSOs (Gibson et al. 2009b), and the X-ray weakness increases going from NAL to mini-BAL to BAL QSOs (Chartas et al. 2009); the relation could again be either geometrical, with BAL, mini-BAL, and NAL structures coexisting in the same QSO wind and occupying higher and higher latitudes above the accretion disk plane (e.g. Hamann et al. 2012), or evolutive, with e.g. mini-BALs being seeds capable to evolve in BALs if the QSO spectral energy distribution changes favourably (e.g. Gibson et al. 2009b).
The situation is complex and interesting: in recent years there have been observations of non-BAL QSOs developing BALs (Hamann et al. 2008; Krongold, Binette \& Hern\'andez-Ibarra 2010), of NALs evolving into BALs (Ma 2002); of mini-BALs evolving into BALs (Rodr\'iguez-Hidalgo et al. 2013); even the canonical Seyfert 1 NGC 5548 has been recently observed in an X-ray obscured state, and has discovered to have concurrently developed UV mini-BALs (Kaastra et al. 2014).

While the width of mini-BALs is large enough to make them easily identifiable in the optical/UV spectra of moderate-to-high resolution, and to be assumed to be intrinsic to the QSO, determining the intrinsic nature of NAL features is much more difficult, as their narrowness makes them very hard to be distinguished from extrinsic, intervening absorbers.
For these reasons, only recently the first systematic studies have started (e.g. Ganguly et al. 1999; Misawa et al. 2007), only a low number of intrinsic NAL systems is known, and an even smaller number of X-ray observations do exist (see Chartas et al. 2009).
Being the number statistics still very low, this work will focus on mini-BAL and BAL QSOs X-ray/UV properties, leaving the incorporation to the study of the NAL QSOs properties to a future work.

We present here the results of the analysis of XMM-Newton pointed observations of a small sample of BAL and mini-BAL QSOs, bright enough to be studied with moderate resolution X-ray spectroscopy, with the aim to present the flavour of their X-ray spectral characteristics and compare them to the expectations of the different scenarios outlined above. The sample and the observations are introduced in Section~\ref{Sec:2}, data analysis is presented in Section~\ref{Sec:3}, and Conclusions are drawn in Section~\ref{Sec:5}. 
A cosmology with $H_0 = 70$ km s$^{-1}$ Mpc$^{-1}$, $q_0=0$, and $\Omega_{\Lambda}=0.73$ is adopted throughout the paper.
Statistical errors are quoted at $1\sigma$ confidence level if not otherwise stated. 

\section{Sample and Observations\label{Sec:2}}
 
A two-fold strategy has been pursued in the last years in order to overcome the X- ray weakness of BAL QSOs and determine their high-energy properties. A hardness ratio and/or stacking analysis technique was applied to snapshot observations of a large number of sources in order to characterise their mean properties (e.g. Green et al. 2001; Gallagher et al. 2006), while detailed spectral analysis of deep observations was only possible for a few bright sources with pointed observation of either ROSAT, BeppoSAX, ASCA, Chandra, or XMM-Newton, and have often revealed interesting complexities in the absorbers: ionised, blueshifted, partial covering, and variable X-ray absorption have all been detected at least once, at least in one BAL or mini-BAL QSO. However, such studies have been sporadic and related to sources selected on the base of their (relatively) high X-ray flux (e.g. Gallagher et al. 1999; Wang et al. 1999; Gallagher et al. 2002; Chartas et al. 2002; Grupe, Mathur \& Elvis al. 2003; Gallagher et al. 2004; Shemmer et al. 2005; Miller et al. 2006; Schartel et al. 2007; Ballo et al. 2008; Schartel et al. 2010; Giustini et al. 2011; Ballo et al. 2011; Scott et al. 2015).  

Saez et al. (2012) started to unite the two strategies by spectroscopically studying a sample of $11$ BAL QSOs with available multiple X-ray observations in order to search for spectral variability. 
We follow here a similar approach, with the additional requirements of drawing the sample from a single catalogue, and of using X-ray data taken with a single instrument.
Our aim is to study a small but sizeable sample of BAL and mini-BAL QSOs with moderate quality, possibly time-resolved X-ray spectroscopy, and discuss the results in the context of the accretion disk wind scenarios for AGN. 
To this end, we selected from the Palomar-Green catalog (PG QSOs; Schmidt \& Green 1983) all the QSOs known to host either BALs or mini-BALs, with XMM-Newton pointed observations publicly available as of April 2015. This resulted in $13$ sources with a  total of $46$ XMM-Newton pointed observations.
 Source and observations properties are reported in Tables~\ref{Table1} and \ref{Table2}.
 
 \begin{table}
\caption{Sample Properties\label{Table1}}
\centering
\scriptsize{\begin{tabular}{ccccccc}
\hline\hline
Name &  RA           & dec & $z$ & $N_H^{Gal}$ & Type & Ref \\
(1) & (2) & (3) & (4) & (5) & (6) &(7) \\
\hline
PG 0043+039 & 00 45 47.2 & +04 10 23 & 0.385  & 3.25 & B & a \\  
PG 0935+417 & 09 38 57.0 & +41 28 21 & 1.957  & 1.42 & mB & \\
PG 1001+054 & 10 04 20.1 & +05 13 00 & 0.161 & 1.83 & B & b\\
PG 1004+130 & 10 07 26.1 & +12 48 56 & 0.241 & 3.56 & B & c,d\\
PG 1114+445 & 11 17 06.4 & +44 13 33 & 0.144 & 1.77 & mB & \\
PG 1115+080 & 11 18 16.9 & +07 45 58 & 1.735 & 3.57 & mB & e,f\\
PG 1126-041 & 11 29 16.6 & -04 24 08 & 0.062 & 4.35 & mB & g\\
PG 1351+640 & 13 53 15.8 & +63 45 46 & 0.088 & 1.98 & mB & \\
PG 1411+442 & 14 13 48.3 & +44 00 14 & 0.089 & 0.87 & mB & \\
PG 1416-129 & 14 19 03.8 & -13 10 45 & 0.129 & 7.08 & B & \\ 
PG 1535+547 & 15 36 38.3 & +54 33 33 & 0.039 & 1.40 & mB & h\\
PG 1700+518 & 17 01 24.8 & +51 49 20 & 0.292 & 2.26 & B & i\\
PG 2112+059 & 21 14 52.6 & +06 07 42 & 0.466 & 6.09 & B & j,k\\
\hline
\end{tabular}\\}
\footnotesize{\textbf{\ \\Notes:} (1) Source name; (2) Right Ascension (J2000); (3) Declination (J2000); (4) Cosmological redshift; (5) Galactic column density in units of $10^{20}$ cm$^{-2}$; (6) Type of source: BAL QSO (B) or mini-BAL QSO (mB); (7) References for previous detailed studies in the literature involving the same X-ray data used here:
a) Kollatschny et al. 2015; b) Saez et al. 2012; c) Miller et al. 2006; d) Scott et al. 2015; e) Chartas, Brandt \& Gallagher 2003; f) Chartas et al. 2007; g) Giustini et al. 2011; h) Ballo et al. 2008; i) Ballo et al. 2011; j) Schartel el al. 2007; k) Schartel et al. 2010. }
\end{table}

The PG catalog is the archetypal sample of UV-excess selected, bright (m$_B < 16$), blue ($U - B < -0.44$) QSOs, and one of the best studied AGN sample with an impressive multiwavelength coverage performed over the past three decades (e.g. Tananbaum et al. 1986; Neugebauer et al. 1987;  Kellermann et al. 1989; Boroson \& Green 1992; Laor et al. 1997; Brandt et al. 2000; Piconcelli et al. 2005; Dasyra et al. 2007). Because of the magnitude and color selection, the PG QSOs are mostly at redshift $z<2$: the average redshift of the sample of mini-BAL and BAL PG QSOs studied here is $\langle z \rangle = 0.445$, with a large scatter - from the lowest redshift PG 1535+547 ($z=0.0389$) to the highest redshift PG 0935+417 ($z=1.957$). Except for the aforementioned blue-ness and brightness, the PG sample properties are well representative of those of the much larger (both in covered area and in redshift) sample given by the SDSS QSOs (Jester et al. 2005). 
The sources of our sample are all radio-quiet with the exception of PG 1004+130; six sources are classified as BAL QSOs, seven as mini-BAL QSOs.

The EPIC-pn (Str\"uder et al. 2001) onboard XMM-Newton offers  the largest effective area among all the flying X-ray detectors, and it is therefore ideal to study X-ray weak sources such as BAL and mini-BAL QSOs.
Data were retrieved from the XMM-Newton Science Archive and were processed using SAS v.14.0 and the most recent calibration files generated in April 2015.
The task \texttt{epproc} was used to concatenate the raw event files and generate calibrated event tables. Good time intervals were selected by discarding periods of strong flaring background, using a uniform threshold of $0.6$ ct s$^{-1}$ applied to a light curve of single events with energies $10$ keV $ < E < 12$ keV and retaining the full field of view.  Source spectra were extracted from circular regions with radii determined with the task \texttt{eregionanalyse}, and span a range $10-54''$. Background spectra were extracted from much larger circular regions, with typical radii of the order of $80''$, in the CCD opposite to the source's one, and avoiding any other evident source of X-ray emission; they were then normalised to the source area with the \texttt{backscale} task. Both single and double pattern events with the highest quality flag (FLAG==0) were retained. For each exposure appropriate response matrices at the source position were generated with the \texttt{rmfgen} and \texttt{arfgen} tasks. The \texttt{specgroup} task was used to group the spectra in order to have a minimum number of 20 counts in each energy bin, allowing the use of the $\chi^2$ statistics during the spectral fitting.

\section{Data analysis results\label{Sec:3}}
Data analysis was performed using \textsc{Heasoft v.6.16} and \textsc{Xspec v.12.9.0d}.
The sources were all detected except for the first observation of PG 0043+039; this source is detected in the second observation performed eight years apart, but the number of counts is too low to allow for a proper spectral fitting. PG 1700+518 was detected in all the three observations, but only at $E<2$ keV. 
Except for these two sources, spectral analysis of the time averaged spectra was performed for each observation of each source applying first simple power law models, then more complex models. Each model included the galactic absorption along the line of sight to the source, as estimated from Kalberla et al. (2005).

First, a simple power law model was fitted to the spectra in the whole $0.2-10$ keV band. The power law fit was then repeated in several different rest frame energy intervals, in order to characterise the broad band spectral shape of all the sources detected with a high enough S/N ($500$ counts in the $0.2-10$ keV band, namely all but PG 0043+039, PG 0935+547, and PG 1700+518). The bands used were $0.2-2$ keV in the observed frame, and $2-5$ and $5-10$ keV band in the rest frame of the sources. 
These bands have been chosen as following: 
\begin{itemize}
\item $ 0.2-2$ keV: approximately corresponds to the ROSAT bandpass. Several low redshift BAL/mini-BAL QSOs were first observed by ROSAT to have a steep power law emission ($\Gamma_{0.2-2}\sim 2.5$), and have been compared to NLS1s, suggesting a common physical scenario (e.g., a high accretion rate, Brandt \& Gallagher 2000). Comparing the slope measured in the $0.2-2$ keV band with the slope of a power law measured in the broad band can assess whether BAL and mini-BAL QSOs really do have a steep intrinsic continuum, or if this is a selection effect due to the narrow energetic band of ROSAT coupled to a complex spectral shape.\\
\item $2-5$ keV: this is the spectral range where AGN are often considered to show their intrinsic continuum, and hence it is often used as the reference spectral range to determine the slope and the amplitude of the power law emission. This is because the strongest reprocessing features in AGN X-ray spectra, absorption and reflection, mostly manifest themselves at energies lower than $2$ keV, and higher than $5$ keV, respectively.\\
\item $5-10$ keV: this is the energy range where both highly ionised absorption and reflection mostly affect the spectral shape, with a steepening of the observed $\Gamma_{5-10}$ the former, with a flattening the latter. A measure of the slope of a power law in this band, compared with the slope of a power law in the $2-5$ keV band, can provide hints on the likely relative contribution of iron K absorption and of reflection to the observed spectral shape. In particular, one can expect $\Gamma_{5-10} >\Gamma_{2-5}$ in the case of strong ionised absorption, vice versa in the case of strong reflection.
\end{itemize}

Most of the sources show complex spectral shapes, as shown by the distribution of the observed photon indices, shown in the top left corner of Figure 1. The vast majority of the spectra significantly deviate from a simple power law emission typical of type 1 AGN, which show $\langle\Gamma \rangle \sim 1.8$ with a dispersion of $\sim 0.2$ (e.g., Piconcelli et al. 2005). In particular, we have $\langle\Gamma\rangle\sim 1.3$ and most of the spectra show a very flat $\Gamma < 1.6$,  suggestive of strong reprocessing in terms of either absorption or reflection. 
This flatness is persistent in the fit performed in the rest-frame $2-5$ keV band, indicating again strong complexity in an energy band that in the average AGN is usually free of strong reprocessing features. The spectral flatness is however not seen in the fit performed in the rest frame $5-10$ keV band alone: it is found $\Gamma_{5-10} > 1.6$ for almost all the spectra. In particular, the fact that $\Gamma_{5-10} > \Gamma_{2-5}$ for the vast majority of the sample is consistent with such spectral complexities to be driven by absorption rather than by strong gravity effects related to relativistic reflection close to the central SMBH.
 As for the fit in the $0.2-2$ keV soft band, almost all the spectra show $\Gamma_{0.2-2}>2$, even if they show $\Gamma < 1 $ when fitted over the whole $0.2-10$ keV band. Of the few spectra with $\Gamma_{0.2-2}<2$, four have typical photon indices, while two are flat, suggesting the possible presence of low ionisation or neutral absorption.
   \begin{figure}
   \centering
  \includegraphics[width=8. cm]{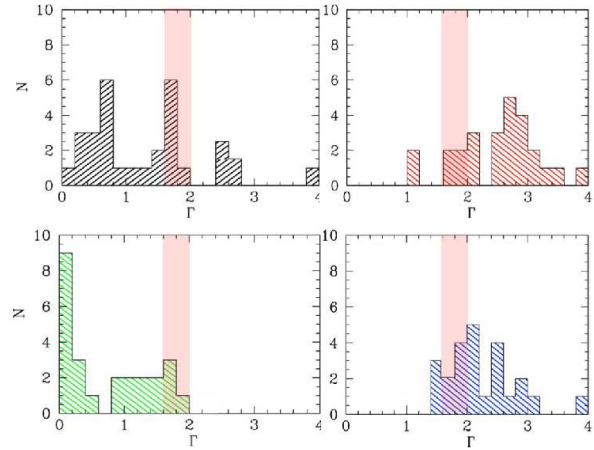}
   \caption{Distribution of photon indices measured for power law fit to the $0.2-10$ keV (top left), $0.2-2$ keV (top right), $2-5$ keV (bottom left) and $5-10$ keV (bottom right) band.}
              \label{FIG:1}%
    \end{figure}

We then tested the presence of absorption in excess of the galactic one. 
A simple neutral absorber is detected in only one source, PG 1115+080, with a low column density $N_H \sim (1 - 3) \times 10^{21}$ cm$^{-2}$ and variable over year time scales.
The situation completely changes if one allows for the absorber to be only partially covering the continuum source or to be ionised.
For all but two sources analysed here the fit statistics significantly improves by allowing the neutral absorber to be only partially covering the source\footnote{The two sources for which the statistical improvement is not significant with respect to a totally covering neutral absorber are PG 1115+080 and PG 1004+130; however the neutral partial covering model fits well also the spectra of these two sources.} and this model is acceptable (reduced $\chi^2 \lesssim 1.2$) for about a half of the spectra. 
The absorbing column densities derived using this model are high, from $N_H \sim 10^{22}$ cm$^{-2}$ up to several $10^{23}$ cm$^{-2}$ (left panel of Figure 2). 
 The vast majority of the spectra show a covering fraction $C_f > 0.8$, with a peak at $C_f\sim 0.95$; a few spectra show much lower covering fractions scattered around $C_f =0.5$. The two different distributions of $C_f$ correspond to very different spectral shapes, and could reflect different physical scenarios for the sources: for example, in the high $C_f$ scenarios the low fraction of emerging continuum could correspond to a secondary component emerging only at soft energies, such as a scattered component, or an underlying thermal emission due to collisionally ionised gas. 
 With the moderate CCD spectral resolution and the generally low S/N ratio of these spectra, disentangling between these different scenarios with a single-epoch spectrum is not possible. In the spectra where $0.3 < C_f < 0.6$ the effect of the partial covering absorber is greatest around $2-3$ keV and this can not be mimicked by a steep secondary soft component - anyway it could be equivalent, for example, to the effect of a totally covering, ionised absorber and the two would be again difficult to disentangle.
 The intrinsic photon indices are much steeper than in the simple power law scenario, with $\langle\Gamma\rangle\sim 2.3$.
 
    \begin{figure}
   \centering
  \includegraphics[width=8. cm]{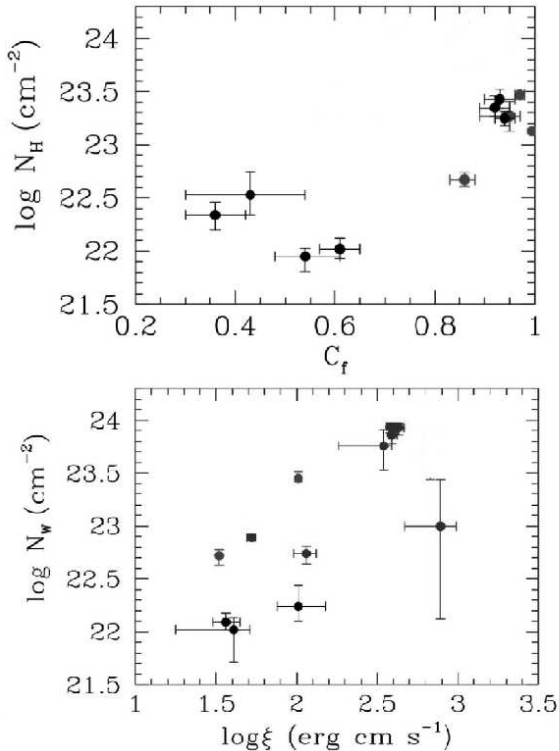}
   \caption{Top panel: covering fraction versus column density in the neutral partial covering scenario. Bottom panel: ionisation state versus column density in the ionised absorption scenario. Errors at $90$\% confidence level.}
              \label{FIG:2}%
    \end{figure}

The presence of ionised absorption intrinsic to the source was also tested by replacing the neutral partial covering absorber with a fully covering ionised one.
It was found that also this model can provide statistically acceptable fits to the majority of the spectra. 
The inferred column densities for this scenario are quite large, with values $ 3-5$ times higher than in the partial covering scenario; this model fits well about a half of the spectra, for which column density versus ionisation parameter are reported in the right panel of Figure 2.
Once corrected for the absorption, most of the sources show photon indices typical of type 1 AGN, with an average $\langle\Gamma\rangle\sim 1.9$.

All the sources with multiple X-ray exposures show strong spectral variability, on different time scales of years, months, days. Furthermore, a number of them also shows variability over very short time scales (a few ks). Given the low S/N ratio and the strong complexity of most of the spectra, a thorough analysis of the timing properties of all the sources of the sample is delicate, and is deserved to a future work. 
We note however how in two of the best S/N spectra it was possible to detect highly ionised outflowing X-ray absorbers in the iron K band, namely PG 1115+080 (Chartas, Brandt \& Gallagher 2003, Chartas et al. 2007) and PG 1126-041 (Giustini et al. 2011). These absorbers are found to be variable over time scales of years (PG 1115+080) and over very short time scales of hours (PG 1126-041).

\section{Conclusions\label{Sec:5}}

The most notable result of the analysis is the extremely complex spectral shape of the vast majority of the BAL and mini-BAL QSOs studied here, coupled to a strong spectral variability on different time scales: years, days, hours, depending on the sources and on the available data. The $0.2-2$ keV photon indices are generally very steep, as already revealed by ROSAT. However, this is the result of a complex spectral shape coupled with the limited ROSAT bandpass: when the larger bandpass of EPIC-pn is considered, the photon indices flattens considerably for all the sources. Almost none of the spectra require the presence of intrinsic neutral absorption when modelled with a simple absorbed power law. On the other hand, high column densities ($N_H \sim 10^{22-24}$ cm$^{-2}$) of X-ray absorbing gas are required by the data
of basically all the sources, when allowing the absorber to be either partial covering or ionised. 
However, only about a half of the spectra are statistically well reproduced by either a neutral partial covering or a ionised absorber model alone. 
The other half of the spectra resulted to be more complex, and the limited S/N ratio makes testing more complex models very difficult. 
Also for the simple partial covering and ionised absorber models there is a substantial degeneracy and they provide similar fit statistics, making it hard to discriminate among the two.
 A lot of the sources show however the clear sign of high column densities of ionised gas affecting their X-ray spectra, as deep absorption troughs around $0.5-1.5$ keV in their rest-frame, an energy range where the opacity of warm absorber-like gas is maximum. 
 Much more highly ionised gas such as Fe XXV and Fe XXVI could also be affecting the spectra of several sources at high energies, given the residuals often visible at $E\gtrsim 6$ keV. 
 This possibility is also suggested by the steep photon index $\Gamma_{5-10}$ found for the vast majority of the sources, and generally also with $\Gamma_{5-10}>\Gamma_{2-5}$. 
 Indeed, highly ionised iron is surely present in the inner regions of the mini-BAL QSOs PG 1115+080 and PG 1126-041.  
 The spectral variability between different exposures is dramatic in several cases, especially at $E\lesssim 6$ keV. 
 
 All in all, these observational results are compatible with the presence of either radiation- or magnetically-driven accretion disk winds originating in the inner regions of AGN. Large column densities of X-ray absorbing gas (perhaps an inner ``failed wind'') are shielding the UV-absorbing photons from the strong continuum emission, preventing them from becoming over ionised, and therefore further preventing them from losing all the resonant line pressure driving force in the case of radiation-driven winds. 
 
Spectral variability is an important piece of information about the accretion/ejection flow in AGN. 
The shortest time scale variability, in particular, can unveil the flow in the very inner regions around the SMBH and discriminate against different launching/accelerating mechanisms, therefore clarifying both the physical structure of the inner regions of AGN and the total contribution of disk winds to the feedback in the surrounding environment. 
 In the sources where spectral variability was observed on ks time scales (notably PG 1114+445, PG 1126-041, PG 1535+547, and PG 2112+029), future deep X-ray observations with high-resolution instruments such as the SXS onboard ASTRO-H (Takahashi et al. 2012) or the X-IFU onboard ATHENA (Barcons et al. 2015) could start to effectively map the dynamics of the inner accretion/ejection flow around supermassive black holes.

\acknowledgements
MG would like to thank Massimo Cappi, George Chartas, Daniel Proga, Gabriele Ponti, Cristian Vignali, Mike Eracleous, and Giorgio G. C. Palumbo for thoughtful discussions and significant inspiration over many fundamental years.
SRON is supported financially by NWO, the Netherlands Organisation for Scientific Research.
Based on observations obtained with XMM-Newton, an ESA scientific mission with instruments and contributions directly funded by ESA Member States and NASA.
This research has made use of the NASA/IPAC Extragalactic Database (NED) which is operated by the Jet Propulsion Laboratory, California Institute of Technology, under contract with the National Aeronautics and Space Administration, and of the NASA's Astrophysics Data System.


\begin{table*}
\caption{X-ray observations log\label{Table2}}
\centering
\small{\begin{tabular}{cccccccccc}
\hline\hline
Name & Date & OBSID & mode & Filter & $t_{exp}$ & $t_{net}$ & $r$ & c-rate & $f_{2-10}$ \\
(1) & (2) & (3) & (4) & (5) & (6) &(7) & (8) & (9) & (10)  \\
PG 0043+039 & 2005-06-15 & 0300890101 & FF & th & $31.8$ & $22.2$ &  - & - & - \\
                       & 2013-07-18 & 0690830201 & FF & th & $33.0$ & $19.4$ & $10$ & $0.0029\pm{0.0004}$ &$1.4\pm{0.4}\times 10^{-14}$  \\
PG 0935+417 & 2007-05-06 & 0504621001 & FF & th & $20.5 $& $7.0 $& $13$ & $0.010\pm{0.002}$ & $4.7^{+1.6}_{-1.3}\times 10^{-14}$ \\
PG 1001+054 & 2003-05-04 & 0150610101 & FF & th & $23.6 $& $10.0 $& $24$ & $0.051\pm{0.002}$ & $1.2^{+1.2}_{-0.6}\times 10^{-13}$\\
PG 1004+130 & 2003-05-04 & 0140550601 & FF & m & $22.2 $& $20.1 $& $33$ & $0.109\pm{0.003}$ & $3.1\pm{0.1}\times 10^{-13}$\\
                       & 2013-11-05 & 0728980201 & FF & m & $66.0 $& $61.4 $& $25$ & $0.068\pm{0.001}$ & $2.8\pm{0.1}\times 10^{-13}$\\
PG 1114+445 & 2002-05-14 & 0109080801 & LW & th & $43.5 $& $38.3 $& $50 $& $0.752\pm{0.005}$& $2.25^{+0.09}_{-0.08}\times 10^{-12}$ \\
                       & 2010-05-19 & 0651330101 & LW & th & $37.8$ & $25.7$ & $34$ & $0.370\pm{0.004}$& $1.76^{+0.10}_{-0.09}\times 10^{-12}$ \\
                       & 2010-05-21 & 0651330201 & LW & th & $34.3$ & $2.8 $& $32$ & $0.43\pm{0.01}$& $2.0^{+0.4}_{-0.2}\times 10^{-12}$\\
                       & 2010-05-23 & 0651330301 & LW & th & $35.8$ & $6.1 $& $38 $& $0.354\pm{0.008}$& $1.7^{+0.2}_{-0.3}\times 10^{-12}$ \\
                       & 2010-06-10 & 0651330401 & LW & th & $41.4$ & $12.2$ & $39$ & $0.541\pm{0.007}$& $2.0\pm{0.1}\times 10^{-12}$ \\
                       & 2010-06-14 & 0651330501 & LW & th & $37.1$ & $6.2 $& $38 $& $0.496\pm{0.009}$& $2.1\pm{0.2}\times 10^{-12}$ \\
                       & 2010-11-08 & 0651330601 & LW & th & $34.3$ & $19.9 $& $53$ & $0.984\pm{0.007}$& $3.3\pm{0.1}\times 10^{-12}$ \\
                       & 2010-11-16 & 0651330701 & LW & th & $34.3$ & $18.7 $& $45$ & $0.680\pm{0.006}$&$2.5\pm{0.1}\times 10^{-12}$  \\ 
                       & 2010-11-18 & 0651330801 & LW & th & $34.3$ & $23.7 $& $41$ & $0.528\pm{0.005}$& $2.1\pm{0.1}\times 10^{-12}$ \\
                       & 2010-11-20 & 0651330901 & LW & th & $34.2$ & $24.3 $& $44 $& $0.654\pm{0.006}$& $2.3\pm{0.1}\times 10^{-12}$ \\ 
                       & 2010-11-26 & 0651331001 & LW & th & $28.9 $& $19.9 $& $42 $& $0.550\pm{0.005}$& $2.0\pm{0.1}\times 10^{-12}$ \\
                       & 2010-12-12 & 0651331101 & LW & th & $28.9$ & $15.1 $& $44 $& $0.598\pm{0.007}$& $2.3\pm{0.1}\times 10^{-12}$ \\
PG 1115+080 & 2001-22-25 & 0082340101 & FF & th & $63.2 $& $53.2 $& $32 $& $0.202\pm{0.002}$ & $3.66\pm{0.07}\times 10^{-13}$ \\
                       & 2004-06-10 & 0203560201 & FF & th & $81.9$ & $67.1$ & $35$ & $0.214\pm{0.002}$ & $4.42^{+0.07}_{-0.06}\times 10^{-13}$ \\
                       & 2004-06-26 & 0203560401 & FF & th & $86.5$ & $71.4$ & $29 $& $0.174\pm{0.002}$& $3.78^{+0.07}_{-0.06}\times 10^{-13}$ \\
PG 1126-041 & 2004-12-31 & 0202060201 & LW & th & $33.8$ & $31.0$ & $36$ & $0.199\pm{0.003}$ & $1.1\pm{0.1}\times 10^{-12}$ \\
                      & 2008-06-15 & 0556230701 & FF & m & $31.4$ & $2.1$ & $35 $& $0.31\pm{0.01}$& $1.4\pm{0.4}\times 10^{-12}$ \\
                      & 2008-12-13 & 0556231201 & FF & m & $11.9$ & $3.1 $& $38 $& $0.60\pm{0.01}$& $2.4\pm{0.3}\times 10^{-12}$ \\
                      & 2009-06-21 & 0606150101 & FF & m & $134.3$ & $92.6$ & $30 $& $0.201\pm{0.002}$&$1.0\pm{0.1}\times 10^{-12}$ \\
                      & 2014-06-01 & 0728180201 & FF & m & $35.9$ & $19.1$ & $27 $& $0.106\pm{0.003}$& $0.4\pm{0.1}\times 10^{-12}$ \\
                      & 2014-06-12 & 0728180301 & FF & m & $23.0$ & $20.0$ & $36$ &  $0.220\pm{0.004}$& $1.3\pm{0.2}\times 10^{-12}$ \\
                      & 2014-06-28 & 0728180401 & FF & m & $28.0 $& $24.5 $& $33 $& $0.162\pm{0.003}$&$0.9\pm{0.2}\times 10^{-12}$  \\
PG 1351+640 & 2004-06-23 & 0205390301 & FF & m &$ 50.9$ & $48.6$ &$ 52 $& $0.718\pm{0.004}$& $6.4\pm{0.4}\times 10^{-13}$ \\ 
                       & 2008-06-08 & 0556230101 & FF & th & $28.7$ & $10.5$ & $28 $& $0.113\pm{0.004}$& $3.9^{+0.9}_{-0.8}\times 10^{-13}$ \\
                       & 2008-06-10 & 0556230201 & FF & th & $29.4$ & $15.5$ & $29 $& $0.125\pm{0.003}$& $4.6^{+0.9}_{-0.7}\times 10^{-13}$ \\
PG 1411+442 & 2002-07-10 & 0103660101 & FF & th & $41.8$ & $24.7 $& $34 $& $0.142\pm{0.003}$& $4.6^{+0.8}_{-0.7}\times 10^{-13}$ \\
PG 1416-129 & 2004-07-14 & 0203770201 & LW & th & $49.9$ & $22.5$ & $54 $& $1.447\pm{0.008}$& $2.80\pm{0.02}\times 10^{-12}$ \\
PG 1535+547 & 2002-11-03 & 0150610301 & FF & th & $29.7$ & $13.6 $& $25$ & $0.152\pm{0.004}$& $1.4\pm{0.2}\times 10^{-12}$ \\ 
                       & 2006-01-07 & 0300310301 & FF & th & $27.8$ & $6.9 $& $48 $& $0.63\pm{0.01}$& $2.9\pm{0.3}\times 10^{-12}$ \\
                       & 2006-01-22 & 0300310401 & FF & th & $29.5$ & $18.1$ & $34$ & $0.378\pm{0.005}$& $2.3\pm{0.2}\times 10^{-12}$ \\
                       & 2006-01-24 & 0300310501 & FF & th & $25.9$ & $19.0$ & $36$& $0.432\pm{0.005}$& $2.4\pm{0.2}\times 10^{-12}$\\
PG 1700+518 & 2009-12-19 & 0601870101 & FF & th & $19.4$ & $6.2 $& $10 $& $0.004\pm{0.001}$& -  \\
                       & 2009-12-31 & 0601870201 & FF & th & $19.9$ & $10.3 $& $12 $& $0.006\pm{0.001}$& - \\
                       & 2010-01-02 & 0601870301 & FF & th & $19.9$ & $19.0$ & $10$ & $0.004\pm{0.001}$& -  \\
PG 2112+059 & 2003-05-14 & 0150610201 & FF & th & $17.0$ & $8.7$ & $27$ & $0.135\pm{0.004}$& $3.7\pm{0.7}\times 10^{-13}$ \\
                       & 2005-11-20 & 0300310201 & FF & th & $76.1$ & $72.9$ & $15$ & $0.0112\pm{0.0005}$& $7.5\pm{0.2}\times 10^{-14}$ \\
                       & 2007-05-03 & 0500500601 & FF & th & $24.3$ & $22.4 $& $14 $& $0.015\pm{0.001}$& $1.3^{+0.7}_{-0.5}\times 10^{-13}$ \\
                       & 2007-05-19 & 0500500701 & FF & th & $100.3$ & $55.1$ & $16 $& $0.0164\pm{0.0006}$ & $1.2^{+0.5}_{-0.3}\times 10^{-13}$ \\
                       & 2007-05-21 & 0500500801 & FF & th & $99.2$ & $80.7$ & $16 $& $0.0184\pm{0.0006}$& $1.5^{+0.4}_{-0.3}\times 10^{-13}$ \\
                       & 2001-11-05 & 0500500901 & FF & th & $51.9$ & $50.0$ & $16$& $0.0137\pm{0.0006}$& $1.3^{+0.5}_{-0.4}\times 10^{-13}$ \\
                       \hline
\end{tabular}}\\
\footnotesize{\textbf{Notes:} (1) Name of the source;  (2) Date of observation; (3) Observation ID; (4) Epic-pn observing mode: Full Frame (FF) or Large Window (LW); (5) Optical filter used during the observation: thin (th) or medium (m); (6) Nominal exposure time, in kiloseconds; (7) Net exposure time after accounting for strong flaring background, in kiloseconds; (8) Radius used for the source spectrum extraction region, in arcseconds; (9) net count rate in the $0.2-10$ keV band, in counts s$^{-1}$; (10) Observed flux in the $2-10$ keV band in units of erg~s$^{-1}$~cm$^{-2}$.}
\end{table*}

\end{document}